\documentclass[12pt]{iopart}
\usepackage[utf8]{inputenc}

\expandafter\let\csname equation*\endcsname\relax
\expandafter\let\csname endequation*\endcsname\relax
\usepackage{mathtools}
\usepackage{amssymb}
\usepackage{graphicx}
\usepackage[style=authoryear,natbib,backend=biber,%
firstinits=true,sortfirstinits=true,uniquename=false,%
dashed=false,maxnames=4,minnames=1,maxbibnames=15
]{biblatex}

\usepackage{zotero}

\DeclareNameAlias{sortname}{last-first}
\DeclareNameAlias{author}{first-last}
\DeclareNameAlias{editor}{author}
\DeclareNameAlias{translator}{author}

\renewbibmacro{in:}{%
  \ifentrytype{article}{}{\printtext{\bibstring{in}\intitlepunct}}}
\DeclareFieldFormat[article,periodical]{volume}{\mkbibbold{#1}}

\usepackage{xcolor}
\usepackage{hyperref}
\hypersetup{colorlinks,breaklinks,allcolors=blue!60!black}

\usepackage{enumerate}

\begin{document}

\title{Loopholes in Bell Inequality Tests of Local Realism}

\author{Jan-\AA{}ke Larsson}

\address{Institutionen f\"or systemteknik, Link\"opings Universitet,
  581 83 Link\"oping, Sweden} \ead{jan-ake.larsson@liu.se}

\begin{abstract}
  Bell inequalities are intended to show that local realist theories
  cannot describe the world. A local realist theory is one where
  physical properties are defined prior to and independent of
  measurement, and no physical influence can propagate faster than the
  speed of light. Quantum-mechanical predictions for certain
  experiments violate the Bell inequality while a local realist theory
  cannot, and this shows that a local realist theory cannot give those
  quantum-mechanical predictions. However, because of unexpected
  circumstances or ``loopholes'' in available experiment tests, local
  realist theories can reproduce the data from these experiments.
  This paper reviews such loopholes, what effect they have on Bell
  inequality tests, and how to avoid them in experiment. Avoiding all
  these simultaneously in one experiment, usually called a
  ``loophole-free'' or ``definitive'' Bell test, remains an open task,
  but is very important for technological tasks such as
  device-independent security of quantum cryptography, and ultimately
  for our understanding of the world.
\end{abstract}

\maketitle

\tableofcontents
\markboth{Loopholes in Bell Inequality Tests of Local
  Realism}{Loopholes in Bell Inequality Tests of Local Realism}
\newpage

In a Bell inequality test of local realism, the word ``loophole''
refers to circumstances in an experiment that force us to make extra
assumptions for the test to apply.  For comparison, in the English
language the word refers to an ambiguity in the description of a
system, that can be used to circumvent the intent of the system. One
example is a loophole in a system of law, meaning some unintended
and/or unexpected circumstances where the law does not apply, or
situations that it does not cover. Such a loophole in a law can be
used to avoid the law without technically breaking it, one popular
example of this is taxation law.  An older meaning of the word is a
narrow arrow slit in a castle wall where defenders can shoot arrows at
attacking forces. Such loopholes are narrow because it should be
impossible to enter the castle through them, the intent being to force
attackers to enter through the well-defended main gate.

In our case, the well-defended gate is the Bell theorem
\citep{Bell1964}: that local realist models cannot give the
predictions obtained from quantum mechanics. The Bell inequality is
derived under the assumptions of local realism and is violated by
quantum-mechanical predictions, and therefore local realist models
cannot give quantum-mechanical predictions. However, when testing this
in experiment, we are no longer in the simple, clean, ideal setting of
the Bell theorem. There are unintended and/or unexpected circumstances
that opens possibilities for local realism to give the output of the
experiment, circumstances that constitute loopholes in Bell inequality
tests of local realism.

The two most well-known loopholes are the ``locality'' loophole
\citep{Bell1964} and the ``efficiency'' loophole \citep{Pearle1970}.
There are a number of issues that fall roughly under these two labels,
but before we discuss these a brief introduction into Bell inequality
tests is needed, to review the explicit assumptions made in the
inequalities (see \autoref{sec:local-realism} below), and to address
some experimental circumstances that cannot be categorized as locality
or efficiency problems.  After this, we will look into the two
mentioned loopholes and see that they actually are the base of two
classes of loopholes with slightly different effects and scope,
locality in \autoref{sec:locality} and efficiency in
\autoref{sec:efficiency}.  A brief conclusion will follow in
\autoref{sec:conclusions} with recommendations for an experimenter
that wants to perform a loophole-free experimental test.

\section{Violation of Local Realism}
\label{sec:local-realism}

The seminal paper by \citeauthor{Einstein1935} (EPR,
\cite*{Einstein1935}), asks the question ``Can [the]
Quantum-Mechanical Description of Physical Reality Be Considered
Complete?''  In the quantum-mechanical description of a physical
system, the quantities momentum ($P$) and position ($Q$) are not
explicitly included, other than as (generalized) eigenvalues of
non-commuting measurement operators. EPR argue that these physical
quantities must correspond to an element of physical reality, and that
a complete theory should include them in the description. Therefore,
they argue, the quantum-mechanical description of physical reality
cannot be considered complete. \citet{Bell1964} enhances the argument
by adding a statistical test that essentially shows that this
completeness cannot be achieved, if the world is local.

\subsection{The EPR-Bohm-Bell experiment}
\label{sec:epr-bohm-bell}

The EPR setting is as follows: consider a (small) physical system on
which we intend to measure position $Q$ or momentum $P$. The physical
measurement devices associated with these measurements are mutually
exclusive, and furthermore, the quantum-mechanical description for
this physical system tells us that the measurements $Q$ and $P$ do not
commute. The standard way to interpret this is that the system does
not possess the properties of position or momentum, only probabilities
are possible to obtain from quantum mechanics.

\begin{figure}[t]
  \centering
  \includegraphics{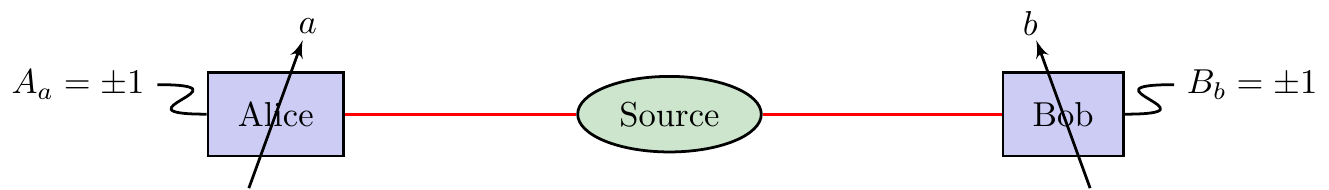}
  \caption{The EPR-Bohm-Bell setup. The two systems are spin-1/2
    systems, and the local measurements are made along a direction in
    space $a$ or $b$, respectively. The source is such that if the
    directions $a=b$, the outcomes $A_a+B_b=0$ with probability one.}
  \label{fig:1}
\end{figure}

EPR now propose using a combined system of two subsystems of the above
type, in a combined state so that measurement of the position sum
gives $Q_1+Q_2=0$ and measurement of the momentum difference gives
$P_1-P_2=0$. These two combinations are jointly measurable, even
though the individual positions and momenta are not, which implies
that it is possible to produce a joint state with these properties.
Letting the two subsystems separate, usually very far, EPR consider
individual measurement of position or momentum. The system is such
that the position sum and momentum difference is preserved under the
separation process, which means that if the position of one subsystem
has been measured, the position of the remote subsystem can be
predicted.  Therefore, EPR argue, the position of the remote subsystem
must exist as a property of that subsystem.  EPR write:
\begin{quote}
  If, without in any way disturbing a system, we can predict with
  certainty (i.e., with a probability equal to unity) the value of a
  physical quantity, then there exists an element of physical reality
  corresponding to this physical quantity.
\end{quote}

Likewise, if the momentum of one subsystem has been measured, the
momentum of the remote subsystem can be predicted. In this case, the
momentum of the remote subsystem must exist as a property of that
subsystem. EPR continue to argue that \emph{both} position and
momentum must simultaneously exist as properties of the remote
subsystem, otherwise
\begin{quote}
  \ldots\ the reality of $P$ and $Q$ [will] depend upon the process of
  measurement carried out on the first system, which does not disturb
  the second [remote] system in any way. No reasonable definition of
  reality could be expected to permit this.
\end{quote}
Bohr's response \citep{Bohr1935} is that
\begin{quote}
  \ldots\ we are not dealing with an incomplete description
  characterized by the arbitrary picking out of different elements of
  physical reality at the cost of sacrifying other such elements, but
  with a rational discrimination between essentially different
  experimental arrangements and procedures which are suited either for
  an unambiguous use of the idea of space location, or for a
  legitimate application of the conservation theorem of momentum.
\end{quote}
Bohr was considered to have won the debate: the consensus at the time
was that there would seem to be no reality to describe. It was not
until much later that Bell realized that Bohr's argument was
incomplete. The exact system that EPR use for their argument
\emph{does} admit a more complete description \citep{Bell1986} but
perhaps, as Bell writes, it was implicitly anticipated already in
\cite*{Bohr1935} that there exist systems that do not. In any case,
the argument needs more support.

The next step forward is the setup proposed in \citeauthor{Bohm1951}
(\cite*{Bohm1951}, see \autoref{fig:1}), that uses a system combined
of two spin-1/2 subsystems. The systems are produced in a total spin 0
state, so that the spins $A_a+B_b=0$ when measured along equal
directions $a=b$.  The subsystems are allowed to separate and a spin
measurement is made on one of the subsystems, and the choice of
measurement directions is a continuous choice instead of the
dichotomic choice in the original EPR setup. Also in this system, the
remote outcome at a given measurement setting can be predicted with
certainty based on the local outcome at that setting.  And because the
reality of the spin measurement outcome in the remote system cannot
depend on the local choice, the spin along \emph{any} direction is an
EPR element of reality. Note that also here, a more complete model
exists for the system than quantum mechanics, when only requiring
prediction with certainty for equal settings at the two sites. But
this complete model does not give the full quantum predictions, a fact
Bell made use of.

\subsection{The Bell inequality}
\label{sec:bell-inequality}

A much stronger and more stringent argument for the non-existence of a
more complete theory than quantum mechanics was provided by J.~S.\
Bell in \cite*{Bell1964}. He took EPR's premise as a starting point
and from the assumptions derived a statistical inequality that is
violated by the quantum-mechanical predictions. This inequality is
derived for EPR-Bohm setup, and must be fulfilled by any mathematical
model that is ``local realist.'' Realism is here motivated by the spin
component along any axis being an EPR element of reality, and locality
is motivated by the finite speed of light, or more specifically,
because local measurement is made ``without in any way disturbing''
the remote system.

In what follows we will use notation from probability theory.  Realist
models (or hidden variable models) are probabilistic models that use
three building blocks. The first building block is a sample space
$\Lambda$, which is the set of possible hidden variable values. The
second is a family of event subsets of $\Lambda$, e.g., the sets of
hidden variable values where specific measurement outcomes occur. The
third and final building block is that these event sets must be
measurable using a probability measure $P$, so that each event has a
well-defined probability. If the sample (hidden variable) $\lambda$ is
reasonably well-behaved, its distribution $\rho$ can be constructed
from $P$. The outcomes of experiment are now described by random
variables (e.g., $A(\lambda)$) that are maps from the sample to the
possible outcomes of the experiment, and the expected (average)
outcome of an experiment can be calculated as
\begin{equation}
  \label{eq:100}
  E(A)=\int_\Lambda A(\lambda)dP(\lambda)=\int_\Lambda A(\lambda)\rho(\lambda)d\lambda,
\end{equation}
the latter if $\rho$ can be constructed. In this notation, the
original Bell inequality can be written as follows.

\smallskip\noindent \hypertarget{thm:1}{\emph{Theorem 1
  \citep{Bell1964}:}} The following four prerequisites are assumed to
hold except at a set of zero probability:
\begin{enumerate}[(i)]
\item \label{B1} \emph{Realism.} Measurement outcomes can be described
  by two families of random variables ($A$ for site 1 with local
  setting $a$, $B$ for site 2 with local setting $b$):
  \begin{equation}
        A(a,b,\lambda)\text{ and }B(a,b,\lambda).
  \end{equation}
  The dependence on the hidden variable $\lambda$ is usually
  suppressed in the notation.
\item \label{B2} \emph{Locality.}  Measurement outcomes are
  independent of the remote setting:
  \begin{equation}
    \begin{split}
      A(a,\lambda)&\stackrel{\textrm{def}}= A(a,b_1,\lambda)= A(a,b_2,\lambda)\\
      B(b,\lambda)&\stackrel{\textrm{def}}= B(a_1,b,\lambda)= B(a_2,b,\lambda).
    \end{split}
  \end{equation}
  For brevity denote $A_i(\lambda)=A(a_i,\lambda)$ and
  $B_j(\lambda)=B(b_j,\lambda)$.
\item \label{B3} \emph{Outcome restriction.}  Measurement outcomes are
  $\pm 1$:
  \begin{equation}
      |A(a,\lambda)|=|B(b,\lambda)|=1
  \end{equation}
\item \label{B4} \emph{Complete anticorrelation.}  At equal settings,
  measurement outcomes are opposite:
  \begin{equation}
   \text{If }a=b\text{ then } A(a,\lambda)=-B(b,\lambda).
  \end{equation}
\end{enumerate}
Then, if $a_1=b_1$,
\begin{equation}
    \Big|E\big(A_2B_1\big)-E\big(A_2B_2\big)\Big|\le 1+E\big(A_1B_2\big).
\end{equation}

\smallskip\noindent \emph{Proof:} Since $a_1=b_1$ we have $B_1A_1=-1$
(with probability 1), and
\begin{align}
  \label{eq:1}
  \Big|E\big(A_2B_1\big)-E\big(A_2B_2\big)\Big|
  &=\Big|E\big(A_2B_1+A_2B_1A_1B_2\big)\Big|\\
  &\le E\Big(\big|A_2B_1\big(1+A_1B_2)\big|\Big)
  = 1+E\big(A_1B_2\big).\tag*{$\Box$}
\end{align}

\subsection{Complete anticorrelation, and determinism vs.\ stochastic
  realism}
\label{sec:determ-vs.-stoch}

Bell used deterministic realism in the sense of EPR elements of
reality: perfect predictability because of the complete
anticorrelation.  However, it is never possible to predict with
certainty in an experiment. There will always be a probability
strictly less than unity that the remote site has the predicted value,
because of experimental imperfections. This will be our first example
of a loophole: the loophole of complete anticorrelation.  The proof in
inequality \eqref{eq:1} relies on perfect anticorrelation, so
\hyperref[B4]{assumption~(\ref*{B4})} must be true, otherwise the
equality does not hold.  It is true that the assumption is explicit
and therefore not ``an unexpected circumstance,'' but one could argue
that it is unexpected that a slightly lowered anticorrelation would
make the proof invalid.  This was realized by \citeauthor{Clauser1969}
(CHSH, \cite*{Clauser1969}) who ``present a generalization of Bell's
theorem which applies to realizable experiments:'' the CHSH
inequality. The removal of the anticorrelation assumption was done in
the CHSH paper while the weaker outcome restriction used below first
appeared in \citet{Bell1971}.

\smallskip\noindent \hypertarget{thm:2}{\emph{Theorem 2
    \citep{Clauser1969,Bell1971}:}} The prerequisites
\hyperlink{B1}{(i)~and (ii) of Theorem 1} and the following third
prerequisite are assumed to hold except at a set of zero probability.
\begin{enumerate}[(iii)]
\item \label{C3} \emph{Outcome restriction.}  Measurement outcomes are
  bounded in absolute value by 1:
  \begin{equation}
      |A(a,\lambda)|\le1 \text{ and }|B(b,\lambda)|\le1.
  \end{equation}
\end{enumerate}
Then,
\begin{equation}
  \Big|E\big(A_1B_1\big)+E\big(A_1B_2\big)\Big|
  +\Big|E\big(A_2B_1\big)-E\big(A_2B_2\big)\Big|
  \le 2.
  \label{eq:2}
\end{equation}

\smallskip\noindent \emph{Proof:}
\begin{align}
  &\Big|E\big(A_1B_1\big)-E\big(A_1B_2\big)\Big|
  =\Big|E\big(A_1B_1\pm A_1B_1A_2B_2-(A_1B_2\pm
  A_1B_1A_2B_2)\big)\Big|\label{eq:3}\\
  &\le E\Big(\big|A_1B_1\big(1 \pm A_2B_2\big)\big|+\big|A_1B_2\big(1\pm
  A_2B_1\big)\big|\Big)
  \le 2\pm\Big(E\big(A_2B_2\big)+E\big(A_2B_1\big)\Big).\tag*{$\Box$}
\end{align}

It was also noted in \citet{Bell1971} that if $A_1B_1=-1$ holds, then
the inequality of \hyperlink{thm:1}{Theorem~1} is a simple corollary
of \hyperlink{thm:2}{Theorem~2}. If instead $E(A_1B_1)$ is larger than
$-1$, \hyperlink{thm:2}{Theorem~2} still holds.  Therefore, there are
no severe effects of the loophole. Taking the loophole into account
changes the inequality, but the change is not large, and there will
still be a violation. Somewhat surprisingly, the CHSH inequality is
actually better than the original Bell inequality, giving a higher
violation for different settings $a_i$ and~$b_j$.  We will adapt this
inequality for other loopholes below but will not obtain higher
violation in any other case. However, it is almost always possible to
derive a useful adapted inequality, as we shall see.

Note that we are not, strictly speaking, using EPR elements of reality
anymore, but the weaker notion of stochastic realism
\citep{Clauser1978} that can be formulated as follows:
\begin{quote}
  If, without in any way disturbing a system, we can predict with
  \emph{high probability} (i.e., with a probability \emph{close} to
  unity) the value of a physical quantity, then there exists an
  element of physical reality corresponding to this physical quantity.
\end{quote}
Here, ``close to unity'' is more a matter of taste; the most commonly
used quantum system has prediction with probability $1/\sqrt2$, see
below.  The question of deterministic versus stochastic realism is
important and subtle, but we will here use notation and theorems for
deterministic realism without loss of generality: outcome random
variables $A(\lambda,a)$ whose value are determined directly from the
hidden variable $\lambda$ rather than conditional outcome
probabilities $P(A|\lambda,a)$ that may be less than one so that the
outcomes are not completely determined by $\lambda$.  The latter
stochastic model can be described by a probabilistic model that adds
more hidden variables, that must be local, which means that the
obtained inequalities do not change. Another way to say this is that
the assumption of determinism in the theorems is not a loophole, since
stochastic realist models are equivalent to mixtures of deterministic
ones \citep{Fine1982-1}.

\subsection{Violation from quantum mechanics, and simplifying
  assumptions}
\label{sec:viol-from-quant}

The quantum-mechanical predictions violates the CHSH inequality, and
the largest violation occurs for a total spin zero state, that gives
the correlation
\begin{equation}
  \label{eq:4}
  \big\langle A_aB_b\big\rangle=-\cos(\phi_{ab})
\end{equation}
with $\phi_{ab}$ being the angle between the two directions $a$ and
$b$. Choosing the four directions $\pi/4$ apart in a plane in the
order $b_2$, $a_1$, $b_1$, $a_2$, one obtains
\begin{equation}
  \label{eq:5}
  \Big|\big\langle A_1B_1\big\rangle
  +\big\langle A_1B_2\big\rangle\Big|
  +\Big|\big\langle A_2B_1\big\rangle
  -\big\langle A_2B_2\big\rangle \Big| =2\sqrt2,
\end{equation}
and this violates inequality \eqref{eq:3}. Therefore, the
quantum-mechanical predictions for such a state cannot be obtained
from a local realist model.

The predictions of quantum mechanics are rotationally symmetric, and
it has been common to simply assume rotational invariance of the
experimental setup. For example, \citet{Clauser1978} discuss this,
saying that
\begin{quote}
  no harm is done in assuming [symmetry relations], 
  since they are susceptible to experimental verification.
\end{quote}
They proceed to simplify inequality (\ref{eq:2}) into
$\big|3E(\phi)-E(3\phi)\big|\le2$, which is simpler to maximize and to
measure since only one setting is used at the first site (and both at
the second). While it is true that almost ``no harm is done'' when
using this assumption, it is important that the experimental
verification really is performed. Otherwise, the assumption can fail
and give a loophole in the experiment. In a sense, failure of the
assumption is a systematic error, and the size of this error needs to
be controlled.  Even a good (but non-ideal) experiment will have small
deviations, and the size of these deviations must be taken into
account when calculating the size of the violation. Checking the size
of the errors will need to be done by measuring at the settings that
were eliminated by the assumption, so it seems better to avoid
symmetry assumptions that can be avoided, even if they are relatively
harmless.

Another simplification that can be made is to measure the outcomes
from the two subsystems jointly: directly measure the product of
$A_aB_b$ rather than the individual $A_a$ and $B_b$. But this enables
a loophole of ``individual existence:'' the individual values must be
assumed to exist since they are not realized in experiment. It is no
longer possible to use the EPR argument for their existence, nor make
predictions to the remote site, since one does not have the local
value to make predictions from. It is not even possible to use quantum
mechanics to argue for the individual existence since the individual
measurements are not performed, and ``unperformed experiments have no
results'' in quantum mechanics \citep{Peres1978}.  A restricted
version of this loophole is present in \citet{Rowe2001} who measure
brightness of two ions in an ion trap. They perform a CHSH experiment,
but use joint output values that are three-valued: no bright ions, one
bright ion, or two bright ions. Assuming that the individual outcomes
are realized, these correspond to $(+1,+1)$, $\{(-1,+1)$ or
$(+1,-1)\}$, and $(-1,-1)$. The obtained correlations do violate the
CHSH inequality, but with the loophole of individual existence
\citep{Danforth2014}. In foundational experiments, this loophole can
and should be avoided by measuring the individual outcomes, whose
(local) realism the experiment attempts to disprove. (The experiment
by \citeauthor{Rowe2001} is important from the point of view of other
loopholes, and we will return to it in \autoref{sec:efficiency}.)

Finally, accidental detections at the measurement sites are sometimes
handled as a systematic deviation from the ideal quantum-mechanical
predictions. Photonic experiments in particular are prone to this,
e.g., because of stray light that enters the detectors, or so-called
``dark counts'' --- false detection events that are due to thermal
noise within the detector. These accidentals could now appear to be
valid measurement results, that due to their random nature will
increase the noise in the experiment. An experimenter could now be
tempted to remove this extra noise, by ``subtracting the
accidentals''. It is simple to estimate the size of the extra noise,
and therefore also easy to remove it. But one should be aware that
this opens up a loophole, that the accidentals \emph{and} the higher
correlation that remains after subtracting the accidentals \emph{both
  together} are given by a local realist model. The loophole can be
avoided by making sure that such accidentals occur rarely enough to
give a violation in the raw data, including accidentals. An early
example of such an experiment is \citet{Freedman1972}, who report a
violation after subtracting accidentals, but in a footnote add
\begin{quote}
  In fact, our conclusions are not changed if accidentals are
  neglected entirely; the signal-to-accidental ratio with polarizer
  removed is about 40 to 1 for the data presented.
\end{quote}
This eliminates the loophole; subtraction of accidentals should be
avoided.

\subsection{Experimental violation, and finite sample}
\label{sec:exper-viol-finite}

We now turn to random variations rather than systematic errors.
Because of random variations in experimental data, such data should
not be confused with the quantum-mechanical predictions: experimental
data \emph{can} violate the bound even under local realism. This is
perhaps unexpected, but completely natural because of the
randomness. For example, independent fair coin tosses for the outcomes
gives
\begin{equation}
  \label{eq:6}
  P\big(A_iB_j=\pm 1\big)=\frac12,
\end{equation}
and there certainly exists a local realist model that gives
$E(A_iB_j)=0$, so the predictions do not give a violation.
Nonetheless, the probability is nonzero for an apparent violation in
individual outputs: even with independent fair coins, the prediction
for four independent experiments (superindex 1--4) at the four setting
combinations is
\begin{equation}
  \label{eq:7}
  P\Big(A_1^{(1)}B_1^{(1)}+A_1^{(2)}B_2^{(2)}
  +A_2^{(3)}B_1^{(3)}-A_2^{(4)}B_2^{(4)}=4\Big)=\frac1{16}
\end{equation}
Individual experimental outcomes are simply not bounded by the
inequality.  This is natural because of the random nature of the
experiment, but could be thought of as an unexpected circumstance, a
loophole. The problem here is the finite\footnote[1]{The word
  ``finite'' means non-infinite here; physicists sometimes use the
  word to mean non-zero.} sample size used in experiment, above a
single sample for each setting combination.  A common name for the
problem is ``finite statistics,'' meaning that the sample mean can and
will deviate from the mean of the distribution.

This loophole has also been observed under a different guise, because
of the following observation: The different terms in
\begin{equation}
  A_1^{(1)}B_1^{(1)}+A_1^{(2)}B_2^{(2)}+A_2^{(3)}B_1^{(3)}-A_2^{(4)}B_2^{(4)}
  \label{eq:1000}
\end{equation}
use different values of the hidden variable $\lambda^{(i)}$, while the
CHSH proof uses the same $\lambda$ for all four measurement setting
combinations. Therefore the CHSH inequality does not apply to that sum
of measurement values --- which is completely true, and indeed the
reason for the nonzero probability in equation \eqref{eq:7}. It is in some
sense the reason for the loophole of finite statistics.  However,
although the CHSH inequality does not apply to the above combination
of individual samples, it does apply to the distribution mean.  And
the sample mean will tend to the distribution mean as the sample size
grows; the adherence to the CHSH inequality will become better and
better. To determine to what degree the inequality applies for a given
sample size, we need to use methods from statistics.

Because of the above, obtaining a high value of the CHSH right-hand
side from experiment is not enough. Even a seemingly high violation
could be too low if the sample size is small. This unexpected
circumstance could be thought of as a loophole: the loophole of low
violation. We must determine how strong the evidence is against local
realism, given our finite-sized sample of experimental data.  The CHSH
inequality applies to the correlations, but from experiments we can
only estimate the correlations.  The quality of the estimate is
simplest to judge if we can assume that the outcomes $A_i^{(n)}$ and
$B_j^{(n)}$ are independent identically distributed (IID; this
assumption will give another loophole which we will return to in
\autoref{sec:memory}).  Sometimes this is called ``Poisson
statistics'' although that term is more suitable when looking at a
continuously pumped experiment, where events happen at random times at
a constant rate.  Assuming IID successive outcomes, the standard
unbiased point estimate of the correlation $E(A_iB_j)$ using data from
$N_{ij}$ experimental runs is the sample mean
\begin{equation}
  \label{eq:8}
  E^*\!\big(A_iB_j\big)=\frac1{N_{ij}}\sum_{n=1}^{N_{ij}}A_i^{(n)}B_j^{(n)}.
\end{equation}
Perhaps it is appropriate to point out that this is a random variable,
since it is a sum of random variables. The estimate can now be used on
each of the four terms in the left-hand side of the CHSH inequality to
give
\begin{equation}
  \beta^*=E^*\!\big(A_1B_1\big)+E^*\!\big(A_1B_2\big)
  +E^*\!\big(A_2B_1\big)-E^*\!\big(A_2B_2\big).
\end{equation}
Note that experiments at different setting combinations use different
samples (superindices), as in equation (\ref{eq:7}).  The above is now an
unbiased point estimate of the distribution mean
\begin{equation}
  \beta=E\big(A_1B_1\big)+E\big(A_1B_2\big)
  +E\big(A_2B_1\big)-E\big(A_2B_2\big),
\end{equation}
which is unknown but less than 2 under local realism.

We now need to quantify how close the estimate is to the true mean.
Experimental papers often measure the strength of the test as the
``number of standard deviations'' of violation. One should be aware
that the distribution standard deviation is not known, but the sample
standard deviation is used instead, as the square root of the unbiased
point estimate of $V(A_iB_j)$,
\begin{equation}
  \label{eq:9}
  s_{ij}^2
  =\frac1{N_{ij}-1}\sum_{n=1}^{N_{ij}}\Big(A_i^{(n)}B_j^{(n)}-E^*\!\big(A_iB_j\big)\Big)^2
  =\frac{N_{ij}}{N_{ij}-1}\bigg(1-\Big(E^*\!\big(A_iB_j\big)\Big)^2\bigg).
\end{equation}
The last equality holds here because of the $\pm1$ outcomes of our
experiment. If the independence assumption holds, the variance of the
sum $A_1^{(1)}B_1^{(1)}+A_1^{(2)}B_2^{(2)}
+A_2^{(3)}B_1^{(3)}-A_2^{(4)}B_2^{(4)}$ from equation (\ref{eq:7}) can be
estimated by
\begin{equation}
  \label{eq:10}
  s^2=s_{11}^2+s_{12}^2+s_{21}^2+s_{22}^2.
\end{equation}
An unbiased point estimate for $V\big(E^*(A_iB_j)\big)$ (the variance
of the sample mean) would now be $s_{ij}^2/N_{ij}$, and consequently
an estimate for the variance for $\beta^*$ is
\begin{equation}
  \label{eq:11}
  s_{\beta^*}^2
  =\frac{s_{11}^2}{N_{11}}+\frac{s_{12}^2}{N_{12}}
  +\frac{s_{21}^2}{N_{21}}+\frac{s_{22}^2}{N_{22}}.
\end{equation}
If $N_{ij}$ are close to $N$, then $s_{\beta^*}^2\approx s^2/N$ with
equality if $N_{ij}=N$.  The sample standard deviation for the CHSH
estimate $\beta^*$ is then $s/\sqrt{N}$, which is the usual behaviour
of the estimated standard deviation of the mean as the sample size
grows. Some papers use another approach, and coarse-grain their data
into several decent-sized subsamples, calculate a subsample mean for
each, and use those to calculate $\beta^*$ and $s_{\beta^*}$.  Either
method can be used to obtain the number of standard deviations of
violation
\begin{equation}
k=\frac{\beta^*-2}{s_{\beta^*}}.
\label{eq:12}
\end{equation}

While this is the statistic presented in many papers, a statistician
would prefer to do a hypothesis test. The natural null hypothesis is
local realism, and a statistician would decide on the level of
significance of the test before doing the experiment. The level of
significance is then compared to the probability of obtaining the
observed violation or higher, under local realism. The experiment is a
Binomial trial, but for simplicity here we assume that our sample is
large enough to use the central limit theorem, which allows us to
approximate the distribution of $\beta^*$ with the Normal
distribution. An often quoted number for ``large enough'' in a
Binomial trial is $Np>5$ where $p$ is the smallest probability of one
outcome. The quantum-mechanical smallest probability of $A_iB_j=\pm1$
is $p=(2-\sqrt2)/4$, so that $N\gtrsim35$ would be large enough. Since
we do not know $p$, this only gives a rough idea on when Normal
approximation could be expected to work, the curious should look into
the standard literature to learn more. However, modern experiments are
typically run with $N$ into the thousands. Also, the variance of
$\beta^*$ is not known, so we should use $(\beta^*-\beta)/s_{\beta^*}$
which has the Student $t$-distribution with $N-1$ degrees of freedom,
but again if $N$ is large we can approximate with the Normal
distribution. Note that if the data is coarse-grained as mentioned
above, the number of degrees of freedom is the number of subsamples
minus one, so that the Student $t$-distribution may need to be
used. If Normal approximation can be used and local realism holds so
that $\beta\le2$, then
\begin{equation}
  \label{eq:13}
  P\big(\beta^*\ge 2+ks_{\beta^*}\big)
  \le P\bigg(\frac{\beta^*-\beta}{s_{\beta^*}}\ge k\bigg)
  =1-\Phi\big(k\big)<\frac1{k\sqrt{2\pi}}e^{-\frac{k^2}2},
\end{equation}
(where $\Phi$ is the Normal cumulative distribution function) and if
$N_{ij}=N$ this reads
\begin{equation}
  \label{eq:14}
  P\bigg(\beta^*\ge 2+k\frac{s}{\sqrt{N}}\bigg)
  <\frac1{k\sqrt{2\pi}}e^{-\frac{k^2}2}.
\end{equation}
The number $k$ can be chosen arbitrarily, and can be used to translate
the number of standard deviations of violation into a probability of
getting the observed violation under local realism, essentially
behaving as exp$(-k^2/2)$. This enables a proper hypothesis test,
closing the loophole of finite statistics.

Different Bell tests have different statistical properties. There is
now, some fifty years after Bell's paper, a long list of Bell
inequalities with different properties. Examples include higher
dimension, more sites, and different distribution of entanglement
between the sites. The above analysis can be further refined to
compare the statistical strength of different Bell inequality tests
\citep[van][]{vanDam2005}, but as the authors of that paper point out,
statistical strength is not all, one has to look at how difficult the
experiments are to implement well so that the statistical strength is
useful, and also the rate at which the samples are obtained so that a
large sample can be generated. Currently, it seems that low dimension
and few sites is the most popular alternative.

\section{Locality, memory, and freedom of choice}
\label{sec:locality}

Locality is an explicit assumption, so the term ``locality loophole''
is a slight misnomer. An explicit assumption is not an unexpected
circumstance; also, locality is present in the very name of the
concept local realism and in the ubiquitous term Local Hidden
Variables. There can be no doubt that it is used in any Bell
inequality.  Nonetheless, \emph{failure} of locality might be an
unexpected circumstance, and for that reason we will have a look at a
few variants of the locality loophole, and the reasons why these are
unexpected in the experimental context.

\subsection{Locality and fast switching}
\label{sec:local-fast-switch}

It is certainly true that if communication between the sites is
possible (so that locality is not enforced), then local hidden
variable models are possible.  If measurement settings are chosen and
set long before an experimental run, there is nothing that prevents a
signal from traveling from one site to another, a signal that can
carry influence from the remote setting to the local outcome.  The
importance of changing the measurement settings quickly was first
stressed in \citet{Bohm1957}, in the context of EPR-Bohm experiments,
noting that
\begin{quote}
  \ldots\ particle B (which does not interact with A or with the
  measuring apparatus) realizes its potentiality for a definite spin
  in precisely the same direction as that of A. \ldots\ Such an
  interaction \ldots\ would have to be instantaneous, because the
  orientation of the measuring apparatus could very quickly be
  changed, and the spin of B would have to respond immediately to the
  change.
\end{quote}
\citet{Bell1964}, considers this in conjunction with his inequality,
but not as a loophole in the proof of the inequality. He is well aware
that if locality is not enforced in an experimental test, then
violation does not say anything about properties of the
model. Instead, he wonders if there will be a violation at all if
locality is enforced; perhaps nature will simply deviate from the
quantum-mechanical predictions, allowing an underlying local realist
theory.  After all, at the time the quantum-mechanical predictions
were not known to hold if the settings are chosen and set so that
locality is enforced. While this was certainly expected, already
\citet{Furry1936} had theorized that the quantum wave function would
separate into factors at large distances, which would mean there would
be no violation. An experimental test was still missing, except for
using static settings at parallel and perpendicular directions
(\cite{Wu1950}; see also \cite{Kocher1967}). And, as \citet{Bell1964}
writes:
\begin{quote}
  In that connection, experiments of the type proposed by
  \citeauthor{Bohm1957} [\cite*{Bohm1957}], in which the
  settings are changed during the flight of the particles, are
  crucial.
\end{quote}
In this sense, the loophole is not so much the possibility of a hidden
subluminal communication from one site to the other in slow-changing
measurement setups, but rather that the quantum violation might
disappear in fast-changing setups. This would constitute the
unexpected circumstance, that could cause failure of the conclusion
that local realist theories cannot give the same correlations as found
in nature. The correlation could be low enough to allow local realism
in a fast-changing setup. Then, closing the loophole is not the actual
process of enforcing fast changes of measurement settings, but instead
ensuring that the violation remains while doing so.

It is possible to view the loophole differently: that the possibility
of slower-than-light communication invalidates the inequality.  The
unexpected circumstance would in that case be the unknown mechanism
for communication. An influence is in principle allowed by special
relativity, but it does not correspond to any known force or influence
in any established physical theory.  In this form of the loophole,
prevention of communication is the focus, but of course one should
make sure that the violation remains. Since prevention of
communication is important here, one can ask what correlations a
bounded but non-zero amount of communication would allow, and some
early papers on this are \citet{Brassard1999,Steiner2000,Toner2003}.

The first experiment to close the locality loophole was
\citet{Aspect1982}, which really focuses on the loophole in the first
sense above. This groundbreaking experiment is the first to show that
even with fast-changing settings, the quantum-mechanical predictions
remain, and therefore also the violation of the Bell inequality
remains.  We will continue here to look at some weaknesses of this
experiment, but bear in mind that it is the first experiment that
tests quantum mechanics in a new regime, that rules out suggestions of
the type that \citet{Furry1936} made, and answers Bell's
(\cite*{Bell1964}) worries that the quantum-mechanical predictions
might not hold when the measurement settings change rapidly. Bell was
acutely aware that such an experiment was needed and urged Aspect and
his co-workers to finish the experiment \citep[see e.g.,][]{Bell1981}.
The experiment by \citeauthor{Aspect1982} has had an enormous impact
on the field, and rightly so.

In \citet{Aspect1982}, the distance from each measurement site to the
source is 6 meters, or 20 ns. The settings are switched every 10 ns,
and the coincidence window used is 18 ns, closing the locality
loophole as already stated. It is still vulnerable to the efficiency
loophole (\autoref{sec:efficiency}), but another weakness of the setup
remains: the settings are switched periodically using an ultrasonic
standing wave. Quantum mechanics does not care how the settings are
chosen, but a local realist model might. Two problems arise from this,
the first slightly less important problem \citep{Zeilinger1986} has to
do with an unfortunate property of the geometry of the physical setup
used: twice the period of the switching coincides with the
light-distance between the measurement sites. This means that a
light-speed signal carrying the remote setting from the previous
period would arrive when the local measurement is performed. Avoiding
this problem is simply to vary the period of the switching, or use
properly random measurement settings, see below. The second problem is
that the setting sequence is predictable, and is more serious and
enables a loophole in its own right, the memory loophole.

\subsection{Memory, setting prediction, and independent experiments}
\label{sec:memory}

It was already noted in the proposal of the experiment
\citep{Aspect1976} that a periodic or predictable sequence of settings
is problematic in itself. He notes that if predictable settings are
used,
\begin{quote}
  a supplementary assumption should be exhibited: The polarizers have
  no ``memory,'' i.e., they can be influenced by signals received at a
  certain time from the [remote site] (with a certain delay) but they
  cannot store all this information for a long time and extrapolate
  in the future even if there is some regularity in the [settings].
\end{quote}
In short: a local realist model can give a violation if it can
remember earlier settings and from that predict what the current
remote setting is. The assumption of no memory is needed in the Aspect
experiment, while presence of memory would be an unexpected
circumstance, a loophole.

The assumption would not be needed if properly random or unpredictable
measurement settings are used, and this is done in the experiment by
\citet{Weihs1998}.  There, the measurement setting was applied through
fast electro-optic modulators, so that the setting can be chosen
independently for each experimental run (how to count these is
discussed in \autoref{sec:efficiency}).  It now remains to choose the
settings randomly, and this is done via a physical random number
generator built from a light-emitting diode, a beam-splitter, and two
photomultipliers. Signals from the two photomultipliers are then used
to select the local setting, in fast electronics to ensure locality of
the setting choice.  In this experiment, the distance between the
measurement sites is 400 m or 1.3 $\mu$s, and the total delay from the
random number generation to the measurement has finished is less than
100 ns. The coincidence time window is 6 ns. It is now reasonable to
think that the setting choice is no longer predictable, and then the
assumption of no memory is no longer needed. This setup is often
regarded as having \emph{conclusively} closed the locality loophole.

We shall return to the question of randomness shortly, but there
remains something to be said about memory. Accepting that the settings
are unpredictable, the \citeauthor{Weihs1998} experiment closes the
above memory loophole, because predicting the current remote setting
using the previous sequence of settings is no longer possible.
However, there is another issue here. In the standard statistical
analysis of \autoref{sec:exper-viol-finite}, one assumes that the
experimental runs are independent.  If this is not the case, then one
experiment in principle has access to what settings were used and what
the results were in previously performed experiments, or even the
local hidden variable values. These values might enter into the
process of determining the outcome of the present experiment. Then,
the standard statistical analysis cannot be used, resulting in another
memory loophole \citep{Gill2003,Gill2003-1,Barrett2002}. This
particular memory loophole is not about prediction of the measurement
settings, but rather about dependence of outcomes within a sequence of
experiments.

Removing the loophole requires a statistical analysis that does not
use the independence assumption. If local realism holds, $\beta^*-2$
is a supermartingale \citep{Gill2003-1}, and then a bound can be
obtained for the probability of a large violation of the CHSH
inequality, by using the \citet{Hoeffding1963} inequality.  In the
simplified situation of having equal number of samples in each term of
$\beta^*$ the bound reads
\begin{equation}
  \label{eq:15}
  P\Bigg( \beta^* \ge 2+ k\frac{4}{\sqrt{N}}\Bigg)
  \le e^{-\frac{k^2}2}.
\end{equation}
Again, the constant $k$ can be chosen arbitrarily, just as in the
bound~(\ref{eq:13}).  There is a small correction when the number of
samples differ for different terms in $\beta^*$ \citep{Barrett2002}.
The difference between the bounds \eqref{eq:13} and \eqref{eq:15} is
not so large, the estimated standard deviation $s$ changes to the
number 4 (half the range of expression \eqref{eq:1000}; the quantum
prediction is IID with $s\approx2$), and the factor $1/(k\sqrt{2\pi})$
disappears on the right-hand side. The same bound applies
simultaneously to all earlier $\beta^*$, with less than $N$ samples in
each term \citep[see][]{Gill2003-1}.  This enables a hypothesis test
that closes the memory loophole in a good experiment, meaning a
sizable violation and a large sample.

\subsection{Freedom of choice and superdeterminism}
\label{sec:freed-choice-superd}

So far we have talked about locality of the random variables, that
these are independent of the remote setting. The local result is given
by the hidden variable and the local setting only.  However, to enable
the inequality, it is not enough to have the random variables
independent of the remote setting. The probability measure $P$ (of the
hidden variable $\lambda$) also cannot depend on either local setting
$a$ and $b$.  In a sense, this is more related to our reality
assumption than to the locality assumption. The reality assumption
(\hyperlink{thm:1}{Theorem~1:~i}) is intended to capture that the
model describes an underlying reality independent of measurement.  In
the theorem, the local hidden variable $\lambda$ is a mathematical
representation of the underlying reality, while the settings $a$ and
$b$ are free parameters that are under control of the
experimenter. Formally, the value of $\lambda$ should be independent
of the external parameters $a$ and $b$.

This was discussed in an exchange between
\citet{Bell1976}\nocite{Bell1985}, \citet{Shimony1976} and
\citet{Bell1977}, where the consensus is that this independence is
important. The one difference seems to be stressing that the hidden
variable does not depend on the settings (\citeauthor{Shimony1976}) or
stressing that the settings does not depend on the hidden variable
(\citeauthor{Bell1976}). From a theoretical point of view, the
difference is not so important, because what really matters is that
the hidden variable and the local parameters do not share a common
cause.

In one particular type of experiments, the second kind of independence
is important and actually an essential assumption: so-called passive
choice experiments. In this kind of experiment, one uses the
randomness inherent in the system itself to choose the setting. One
example is, for a photonic experiment, to set up separate detection
stations for each setting at each local site, two at each site for
CHSH, and using a beamsplitter to decide which of the two detection
stations (measurement settings) will be used.  Now, the hidden
variable itself might influence the path through the beam-splitter,
the setting choice, so it is difficult to argue for independence. It
would be possible to outright assume independence in a passive choice
experiment, but this would constitute a loophole in the experiment.

It is argued in \citet{Gisin1999} that when the switches are lossy,
passive switching is almost on equal basis with active switching. The
corresponding assumption for lossy active switching is the fair
sampling assumption (see \autoref{sec:efficiency} below); the
assumption that losses in the switch do not depend on the hidden
variable. But this argument is weak. An active switch has an external
input for an external parameter, the measurement setting choice, which
is an essential part of the Bell Theorem. A passive switch does not
have an input, so the theorem itself does not apply. It is an explicit
assumption that there \emph{is} a random choice that does not depend
on the hidden variable. There is, on the face of it, a big difference
between using the fair sampling assumption (see below) and assuming
that a passive switch is independent of the hidden variable. Both are
assumptions and lead to loopholes, but they are conceptually
different.  The only way to avoid the assumption of passive switch
hidden-variable independence is to use active choice of the
measurement setting.

Now, even in an actively switched experiment one needs to make sure
that measurement settings are chosen independently of the hidden
variables generated at the source. And using locality to argue for
such independence is possible only if the setting choice is made at
space-like separation from the emission event. It is not enough to
have the setting choice inside the future light cone of the emission
event; this only prohibits influence from the setting choice to the
hidden variable, but does not prohibit the reverse influence. Such an
influence would prevent free choice of the measurement settings, and
constitute a ``freedom-of-choice'' loophole, so named in
\citet{Scheidl2010} who report an experiment where the setting choice
is properly space-like separated from the emission event, enabling a
locality-based argument for independence \citep[a three-particle
variation is presented in][]{Erven2014}.  However, the possibility of a
common cause remains, see below.

The question of settings as free parameters is deep; it could be
discussed if it is even in principle possible to have free parameters.
The concept is linked to the concept of free will in the mentioned
exchange between \citeauthor{Bell1976, Shimony1976}, and as
\citet{Bell1977} put it,
\begin{quote}
  Here I would entertain the hypothesis that the experimenters have
  free will.
\end{quote}
For this reason, it is sometimes suggested that an experiment with an
actual human that chooses the experiment settings would be desirable.
But one should be aware of the limitations of such a trial. The sample
rate would need to be low, because a human needs time to produce the
parameter values. Similarly, the distance between source and
measurement site needs to be large, at least on the order of
light-seconds (say, the distance to the moon) for the events to be
space-like separated. Finally, a human subject is typically not a good
source of randomness \citep[see e.g.,][]{Wagenaar1972}, despite having
a free will, or so we believe. While such an experiment would
certainly be interesting, these considerations cast doubt on the added
strength that it would give to the argument against local realism.

Ensuring independence between the source and the settings is a grand
task. One could in principle believe that events that are close in
space depend on each other, because their recent past light-cones
intersect, and that large separation would decrease the dependence.
Already at the conference ``Thinkshops on Physics: Experimental
Quantum Mechanics'' in Erice, Italy, 1976, it was suggested that one
could generate random settings from astronomical objects outside each
others' light cones, e.g., quasars on opposite sides of the universe
\citep{Zeilinger2014}.  The benefit of such an experiment would be to
enlarge the separation of the setting choice.  However, it is
impossible for us to know if the backward light-cones from the two
distant emission events overlap, and if that overlap contains a common
cause for the emissions. The most popular model of the origins of our
universe is as a very small space-time region, so there may still be a
common cause.  Even remote quasar emissions must be \emph{assumed} not
to have a common cause. It is possible that all events in the universe
share common causes, a philosophical view called superdeterminism
\citep[see e.g.,][]{Shimony1976,Bell1977,Brans1988}. This constitutes
a loophole, but if superdeterminism holds, there is no point in
discussing what mathematical models could be used to model nature, be
it local realist or quantum or any other model.  We can never rule out
this possibility using scientific methods, because
\citep{Shimony1976}:
\begin{quote}
  In any scientific experiment in which two or more variables are
  supposed to be randomly selected, one can always conjecture that
  some factor in the overlap of the backward light cones has
  controlled the presumably random choices.  But, we maintain,
  skepticism of this sort will essentially dismiss all results of
  scientific experimentation.  Unless we proceed under the assumption
  that [superdeterminism does not hold,]
  we have abandoned in advance the whole enterprise of discovering the
  laws of nature by experimentation.
\end{quote}
The loophole of superdeterminism cannot be closed by scientific
methods; the assumption that the world is not superdeterministic is
needed to do science in the first place.

\section{Efficiency, coincidence, and postselection: missing events}
\label{sec:efficiency}

We will now turn to a different problem, which ``is a very delicate
one, yet one of great importance'' \citep{Clauser1978}. In an ideal
situation, every experimental run would give registrations in the
measurement devices.  However, this may not be the case in real
experiments. For example, in a photonic Bell experiment, a detection
at one site will not always be accompanied with a detection at the
other.  In fact, it is not always well-defined what an ``experimental
run'' means, since if there are local losses it may happen that both
particles are lost. And there may be no indication of this in the
experiment, so that there is no corresponding event recorded in the
experimental data. This unexpected circumstance makes the original
Bell inequality derivation invalid and is known as the ``detector
efficiency'' or simply the ``efficiency'' loophole.

A few recent experiments do have well-defined experimental runs, and
register outcomes for every experimental run, so that the loophole is
not present. The first such experiment is that of \citet{Rowe2001}, an
experiment on two ions in a trap. In this experiment, every
experimental run gives output data, so it is free of the efficiency
loophole, but the locality loophole is present since the ions are 3
$\mu$m apart while a measurement lasts 1 ms ($\approx$ 300 km). A
later experiment entangles two ions in separate traps at 1 m distance
\citep{Matsukevich2008}, which is much better in three senses: first,
measurement results are recorded individually (see
\autoref{sec:viol-from-quant}); second, it is much easier to argue
that the ions are separate systems since they are held in different
traps, and third, the locality issue is improved by seven orders of
magnitude (distance 1 m and measurement time 50 $\mu$s $\approx$ 15
km). Another experiment, this time in a superconducting system
\citep[the phases of two Josephson junctions,][]{Ansmann2009}, is also
free of the efficiency loophole and improves the locality issue
further by a factor of five (3.1 mm versus 30 ns $\approx$ 9 m).

Finally, \citet{Hofmann2012} implements a ``heralded'' source, by
``entanglement swapping'' \citep{Zukowski1993}, that also gives
well-defined experimental runs. This process entangles two separate
systems and gives a clear signal at the source upon success. In the
reported experiment, two atoms in individual atomic traps 20 m apart
are entangled by photon emission from the atoms and a joint Bell state
measurement at a central ``source'' point. The resulting entangled
two-atom system then always gives outcomes for atomic state
measurements, and this experiment gives a small but sufficient
violation of the CHSH inequality. It is still vulnerable to the
locality loophole because atomic state readout is by fluorescence
(distance 20 m versus 60 ms $\approx$ 18000 km, using the detection
scheme of \cite{Volz2006}), but the group is currently building a
long-distance experiment with faster photoionization detectors that
would close the locality loophole \citep[400 m versus 810 ns $\approx$
240 m using the detection scheme of][]{Henkel2010}.

Closing the locality loophole needs a system that is easy to transport
while keeping entanglement intact. The system of choice is photons,
and photon detectors are inefficient --- more accurately, photon
correlation experiments are inefficient.  To enable a direct photonic
test, we need to re-establish a modified bound that is relevant when
the efficiency is decreased.  An additional problem in some photonic
experiments is to identify which detections belong to the same pair of
particles. A third and final problem in some proposals is that some of
the registered events should not be used in the correlation
calculation. All three problems influence the bound of the CHSH
inequality, or any Bell inequality used. In what follows, we will
consider this loophole in more detail in its different flavors:
efficiency, coincidence, and explicit postselection.

\subsection{The efficiency loophole, and fair sampling}
\label{sec:efficiency-sampling}

The simplest situation to analyze is when local events are missing
from the data record, as already mentioned. The problem here is that
not all values of the hidden variable gives a detection, and that the
obtained correlation is really the conditional correlation
\begin{equation}
  \label{eq:49}
  E\big(A_iB_j\big|A_i\text{ det.\ and }B_j\text{ det.}\big).
\end{equation}
Conditional correlations do not add if the conditioning is over
different subsets, so it is not immediately possible to write
\begin{multline}
  \label{eq:17}
  \qquad E\big(A_1B_1\big|A_1\text{ det.\ and }B_1\text{ det.}\big)+
  E\big(A_1B_2\big|A_1\text{ det.\ and }B_2\text{ det.}\big)\\
  = E\big(A_1B_1+A_1B_2\big|\text{\ldots}\big).\qquad
\end{multline}
This is only visible when explicitly writing out the conditioning on
detection, which is rarely done in practice --- resulting in an
unexpected circumstance, a loophole.  The proof of the Bell inequality
will not go through as is, since it needs addition of the
correlations, but there are two ways of re-establishing (modified)
inequalities. One is to add auxiliary assumptions, which will give
back the original inequality, and the other is to take the missing
events into account in the analysis, which will give an inequality
with a modified bound.

The first paper that discusses this in earnest is \citet{Pearle1970},
where the observation is made that ``data rejection'' enables a
deterministic local hidden-variable model to give the same predictions
as quantum mechanics for the coincident events. This is substantiated
by an explicit model that does just that, so that the conditional
correlations (seemingly) violate the Bell inequality.  However, the
model has a coincidence probability that varies with the angular
difference of the two remote settings, which makes it less appealing
because the quantum-mechanical description predicts constant detection
probability.  Pearle derives a bound for the conditional detection
efficiency for models of the type he uses, and the lowest bound he
finds for his varying efficiency is $2\sqrt2-2\approx 82.84$\%. This
number equals the generic bound for the CHSH inequality, which later
has been derived under increasingly general assumptions, see below.

The CHSH inequality from \hyperlink{thm:2}{Theorem~2} is slightly
earlier (\cite*{Clauser1969}), and is intended for an experimental
setup that uses filters, followed by detectors that assigns $+1$ to
detection and $-1$ to nondetection. The authors note that photonic
experiments at that time with polarization filters would not violate
the CHSH inequality ``because available photoelectric efficiencies are
rather small.'' This is not so much noticing a loophole but simply
that there is no violation since the overwhelming amount of $-1$
outcomes would wash the violation out completely. Low photoelectric
efficiencies is a problem but the preceding polarization filters are
better, so they consider applying the inequality to emergence (here
labeled $+1$) and non-emergence ($-1$) of the photon from the filter
instead.

It now remains to connect emergence to detection, and CHSH
(\cite*{Clauser1969}) do this by assuming that detected photons are a
fair sample of the emerging photons,
\begin{quote}
  that if a pair of photons emerges from [the polarization filters]
  the probability of their joint detection is independent of [the
  settings].
\end{quote}
In the notation used here, the assumption is that there is a local
realist model that describes emergence and non-emergence by $A_i=\pm1$
and $B_j=\pm1$, and that
\begin{equation}
  \label{eq:19}
  P\big(A_i=B_j=+1\cap A_i\text{ det.\ and }B_j\text{ det.}\big)
  =c P\big(A_i=B_j=+1\big),
\end{equation}
where $c$ does not depend on $i$ or $j$.  Since the emergence
correlations on the right-hand side must obey the CHSH inequality, a
similar inequality can now be derived for the detection rates $R$ at
the $+1$ detectors,
\begin{equation}
  \label{eq:20}
  R\big(A_1B_1\big)+R\big(A_1B_2\big)
  +\Big|R\big(A_2B_1\big)-R\big(A_2B_2\big)\Big|
  \le \max_{i,j}\Big(R\big(A_i\big)+R\big(B_j\big)\Big).
\end{equation}
The problem is now reduced to finding sufficiently efficient
polarizers, which is simpler; the paper mentions polarizing
efficiencies of 0.92--0.95, which would give a violation.

The restricted fair sampling assumption used by CHSH applies only to
the detected $A_i=B_j=+1$ events: that these constitute a fair sample
of all the $A_i=B_j=+1$ events.  The fair-sampling assumption as used
today is slightly extended, and applies to all combinations of $\pm1$
outcomes. In short, one uses the same assumption of existence of a
local realist model that describes the ideal outcomes, and in addition
that the probability of joint detection is independent of the
settings, so that
\begin{equation}
  \label{eq:21}
  P\big(A_i=\pm1\cap B_j=\pm1\cap A_i\text{ det.\ and }B_j\text{
    det.}\big)=c P\big(A_i=\pm1\cap B_j=\pm1\big).
\end{equation}
One could think of this as a constant correction to the probability of
emergence from an ideal polarizing beamsplitter, but usually one does
not explicitly refer to the internal workings of the measurement
devices. With this modern variant of the fair-sampling assumption it
is easy to see that
\begin{equation}
  \label{eq:22}
  P\big(A_i\text{ det.\ and }B_j\text{ det.}\big)=c,
\end{equation}
so that
\begin{equation}
  \label{eq:23}
  P\big(A_i=\pm1\cap B_j=\pm1\big|A_i\text{ det.\ and }B_j\text{
    det.}\big)= P\big(A_i=\pm1\cap B_j=\pm1\big).
\end{equation}
Therefore, since the CHSH inequality applies to the underlying local
realist model for the unobserved ideal outcomes, we also have a CHSH
inequality for the detected events that reads
\begin{multline}
  \Big|E\big(A_1B_1\big| A_1\text{ det.\ and }B_1\text{ det.}\big)
  +E\big(A_1B_2\big| A_1\text{ det.\ and }B_2\text{ det.}\big)\Big|\\
  +\Big|E\big(A_2B_1\big| A_2\text{ det.\ and }B_1\text{ det.}\big)
  -E\big(A_2B_2\big| A_2\text{ det.\ and }B_2\text{ det.}\big)\Big|
  \le 2.
\end{multline}
This is violated by the quantum-mechanical predictions as soon as the
efficiency is nonzero.

One should perhaps point out that fair sampling is not equivalent to
constant detector efficiency.  There exist relatively simple local
realist models with constant lowered efficiency that reproduce the
quantum-mechanical predictions
(\cite{Santos1996,Larsson1999-1,Gisin1999-1}; an earlier example with
small deviations is \cite{Marshall1983}).  Of course, these models
typically do not contain predictions on the level of emergence from
the polarizing filters (or beam-splitters). The few that do contain
such predictions do not violate the CHSH inequality at that level,
while the conditional correlations after detection do give a
violation.

\subsection{Non-detections as outcomes, and estimating efficiency}
\label{sec:non-detection-outcomes}

The efficiency problem was again addressed in \citet{Bell1971}, where
he provided a new proof of the CHSH inequality that does not require
the outcomes to be only $\pm1$, but allows them to take values between
$+1$ and $-1$. He uses this to address inefficiency by assigning the
value 0 to the missing outcomes (Bell credits J.~A.\ Crawford for the
suggestion). Note that the value $-1$ as used in \citet{Clauser1969}
is equally suitable, and will give the same correlations and bounds
when the local outcomes are equally probable.

Assignment of values to missing outcomes needs a locality assumption:
that detection is decided locally, and only depends on the hidden
variable and the local setting. This is the natural local
hidden-variable description of the detection process; formally, it is
the assumption that the events (subsets of $\Lambda$) of detection
\begin{equation}
  \Big\{\lambda:A(a_i,b_j\lambda)\text{ det.}\Big\}\text{ and }
  \Big\{\lambda:B(a_i,b_j\lambda)\text{ det.}\Big\}
  \label{eq:50}
\end{equation}
are independent of the remote setting, so that it is meaningful to write
\begin{equation}
  \Big\{\lambda:A_i(\lambda)\text{ det.}\Big\}\text{ and }
  \Big\{\lambda:B_j(\lambda)\text{ det.}\Big\}.
  \label{eq:51}
\end{equation}
If one, like Bell or CHSH, includes the non-detections as events,
there is no conditioning and the CHSH inequality is recovered.
However, lowered efficiency will mean higher probability of the 0 (or
$-1$) ``outcome'' which in turn will mean lower violation, and as CHSH
already had noted, the available equipment at the time would not
violate the inequality.

Value assignment to no-detection events will require the experimenter
to recognize no-detection events, so that the value used can be
assigned.  One way to do this is to change the experimental setup into
a so-called ``event-ready'' (or ``heralded'') setup, in which there is
a clear signal when a pair has been emitted at the source. Then, value
assignment is possible even for non-detection, and direct calculation
of the detection probability can be performed. The need for
event-ready setups is not specifically stated in \citet{Bell1971}
``but is clear from the context'' \citep{Clauser1978}; assigning 0 to
a non-detection requires knowledge that a detection should have
occurred but did not. This point was later stressed in, for example,
\citet{Bell1981}.  An alternative to a heralded source is heralded
detection: to locally certify that the photon is present at the
measurement site before actually inserting the measurement setting
into the measurement device \citep{Ralph2009}; such an experiment has
yet to be done.  Heralded sources, on the other hand, have been built,
one example is the already mentioned atomic source of
\citet{Hofmann2012}. Here, the signal-of-emission needs to be
generated very carefully, so that the violation remains --- in
quantum-mechanical language, the entanglement should not be broken.

Since event-ready experiments are demanding to set up, one alternative
is to use simplifying assumptions on the distribution of
non-detections.  \citet{Garg1987} assume that non-detections occur at
a constant probability
\begin{equation}
  P(A_i\text{ det.})=P(B_j\text{ det.})=\eta,
  \label{eq:24}
\end{equation}
and that the errors are statistically independent,
\begin{equation}
  P(A_i\text{ det.\ and }B_j\text{ det.})
  =P(A_i\text{ det.})P(B_j\text{ det.})=\eta^2.
  \label{eq:25}
\end{equation}
With these assumptions, the CHSH inequality changes into
\begin{multline}
  \label{eq:26}
  \Big|E\big(A_1B_2|A_1\text{ and }B_2\text{ det.}\big)
  +E\big(A_2B_2|A_1\text{ and }B_1\text{ det.}\big)\Big|\\
  +\Big|E\big(A_2B_1|A_2\text{ and }B_1\text{ det.}\big)
  -E\big(A_2B_2|A_2\text{ and }B_2\text{ det.}\big)\Big|
  \le \frac4\eta-2.
\end{multline}
This gives a the generic bound at the same level as the lowest bound
of \citet{Pearle1970}, and gives a violation from the
quantum-mechanical predictions as soon as
$\eta\ge2\sqrt2-2\approx82.84\%$. As a comparison, the approach of
\citet{Bell1971} with the same assumptions gives a bound of
$2^{-1/4}\approx84.09\%$ (however, this number is not given in Bell's
paper).

\citet{Lo1981} recognize the need of avoiding extra assumptions and
derive a new inequality that explicitly includes lowered efficiency
without the symmetry assumptions \eqref{eq:24} and \eqref{eq:25}.
Their inequality is violated by the quantum-mechanical predictions as
soon as the efficiency exceeds 85.97\%, which at the time was the
best-known bound.  But since they avoid the mentioned assumptions, a
different estimate of the efficiency is needed. \citeauthor{Lo1981}
measure the output rate of the source separately, and compare it with
the rate of detection in the full experiment. However, this leaves
room for a local realist model to behave differently in the
source-rate measurement than it does in the violation measurement;
this is a loophole.  For the proposal of \citeauthor{Lo1981}, since
the used system is built up from material particles (quite heavy Na
atoms), it could be argued that there is no room for this kind of
changing behavior. But it is an extra restriction on the local realist
model, motivated by other physical properties of the system than we
normally use in Bell inequality testing. And the experiment seems very
difficult to perform. It would arguably be better to have a generic
method that can avoid the symmetry assumptions, for all kinds of
systems without this type of extra argument.

Fortunately, the symmetry assumptions \eqref{eq:24} and \eqref{eq:25}
can be avoided; they are needed because it is difficult to connect the
probability $P(A_i$ det.\ and $B_j$ det.$)$ to the experimentally
available conditional detection probabilities $P(A_i$ det.$|B_j$
det.$)$ \citep[see e.g.,][]{Burnham1970}.  In \citet{Larsson1998} it
is shown that the conditional probability can be used directly, by
avoiding value assignment to non-detections altogether; the output of
the measurement device will remain undefined.  In this case, the
random variables $A_i$ and $B_j$ are only defined on subsets of the
hidden-variable space, on the subsets where detections occur. Then,
the original CHSH inequality only applies for the joint subset where
all detections would have occurred, where all $A_i$ and $B_j$ have
well-defined values.

In this case, the efficiency $\eta$ must now be estimated differently,
and the appropriate definition is to use the minimum of the
conditional detection probabilities
\begin{equation}
  \label{eq:27}
  \eta=\min_{i,j}\Big\{P\big(A_i\text{ det.}\big|B_j\text{ det.}\big),\
  P\big(B_j\text{ det.}\big|A_i\text{ det.}\big)\Big\}.
\end{equation}
These conditional probabilities are well-defined and easy to estimate
in experiment, and relieve an experimenter from the difficult task of
estimating $P(A_i$~det.$)$ and $P(A_i$ det.\ and $B_j$ det.$)$. It
suffices to estimate the conditional detection probabilities directly
from experimental data and use the lowest value.  Remarkably, this
gives the same inequality as in \eqref{eq:26}, and the bound
$\eta\ge2\sqrt2-2\approx82.84\%$ is valid for the $\eta$ from
equation \eqref{eq:27} without the auxiliary assumptions \eqref{eq:24} and
\eqref{eq:25}.

In all three approaches: event-ready, with, or without symmetry
assumptions, it is the \emph{efficiency of whole experiment} that is
the important number --- the loophole is often called ``the
detector-efficiency loophole,'' but it is important that the number in
question concerns the whole experiment. For comparison, the fair
sampling assumption that was described earlier essentially allows an
experimenter to ignore the \emph{detector} inefficiency, but still
needs the polarizer/beamsplitter efficiency to be included. The one
exception to this is if event-ready analyzer stations are used
\citep{Ralph2009} ensuring that presence of a photon is signaled
before the setting has been input into the station. Then, losses in
transmission before the event-ready indicator can be ignored, while
losses in the analyzer station still needs to be incorporated in the
analysis.

\subsection{Probabilities, not correlations, and the no-enhancement
  assumption}
\label{sec:prob-not-corr}

In \citet{Clauser1974} another approach is presented, to use
probabilities instead of correlations.  In the notation we have
adapted here, the Clauser-Horne (CH) inequality reads
\begin{multline}
  \label{eq:28}
  -1 \le P(A_1=B_1=1)+P(A_1=B_2=1)+P(A_2=B_1=1)
  -P(A_2=B_2=1)\\
  -P(A_1=1)-P(B_1=1)
  \le 0.
\end{multline}
If detection counts are inserted instead of probabilities, this is
sometimes referred to as the Eberhard inequality, having been
rediscovered and found to be very important and useful in
\citet{Eberhard1993}, more on this below. The CH inequality appears to
be different from the CHSH inequality, but the two are equivalent.
Nonetheless, the specific form of the CH inequality makes it useful
for handling the efficiency loophole, because both single and
coincidence probabilities enter directly. Then, it is clear that
conditional probabilities should not be used. And there is no need to
modify the bound for lowered efficiency; instead, a decreased
efficiency will immediately lower the probabilities so that the
violation is reduced. The coincidence terms will decrease faster than
the single-outcome terms, so there will a be violation only for high
efficiency.

\citet{Clauser1974} realized that because of this, the inequality
would not be useful for the available experiments at the time, so they
added a ``no enhancement'' assumption to recover a violation. They
assume that, for a given value of the hidden-variable $\lambda$ in a
stochastic local realist model,
\begin{quote}
  the probability of a count with a polarizer in place is less than or
  equal to the probability with the polarizer removed.
\end{quote}
Denoting a count at site A with the polarizer removed by $A_\infty=1$,
this is the pointwise assumption (in $\lambda$, allowing briefly for
stochastic hidden variables)
\begin{equation}
  \label{eq:29}
  \begin{split}
    P\big(A_i(\lambda)=1\big|\lambda\big) \le
    P\big(A_\infty(\lambda)=1\big|\lambda\big),\\
    P\big(B_j(\lambda)=1\big|\lambda\big) \le
    P\big(B_\infty(\lambda)=1\big|\lambda\big).
  \end{split}
\end{equation}
Under this assumption the inequality changes so that it contains only
coincidence probabilities, some with the polarizers removed:
\begin{multline}
  \label{eq:30}
  -P(A_\infty=B_\infty=1) \le P(A_1=B_1=1)+P(A_1=B_2=1)+P(A_2=B_1=1)\\
  -P(A_2=B_2=1)
  -P(A_1=B_\infty=1)-P(A_\infty=B_1=1)
  \le 0
\end{multline}
This is more restrictive on both sides. The improvement is large when
$P(A_\infty=B_\infty=1)$ is close to $P(A_i$ and $B_j$ det.$)$, i.e.,
when the polarizers have low loss. Similarly to CHSH with non-ideal
polarizing filters and fair sampling, the remaining problem is
polarizer efficiency.  If the polarizer loss is zero (ideal
polarizers), there is a violation as soon as the detection efficiency
is greater than zero, under the no-enhancement assumption.

To illustrate this, a CHSH-like inequality can be derived from
inequality \eqref{eq:30}. If the non-detection events are assigned the
value 0 and we assume no-enhancement, the inequality reads
\begin{equation}
  \label{eq:31}
  \Big|E\big(A_1B_1\big)+E\big(A_1B_2\big)\Big|
  +\Big|E\big(A_2B_1\big)-E\big(A_2B_2\big)\Big|
  \le2P(A_\infty=B_\infty=1).
\end{equation}
The right-hand side is lower than in the usual CHSH inequality, but so
are the correlations to the left because of the lowered efficiency.
Ideal polarizers correspond to $P(A_\infty=B_\infty=1)=P(A_i$ and
$B_j$ det.$)$, and under no-enhancement we arrive at
\begin{multline}
  \label{eq:32}
  \Big|E\big(A_1B_1\big|A_1\text{ and }B_1\text{ det.}\big)
  +E\big(A_1B_2\big|A_1\text{ and }B_2\text{ det.}\big)\Big|\\
  +\Big|E\big(A_2B_1\big|A_2\text{ and }B_1\text{ det.}\big)
  -E\big(A_2B_2\big|A_2\text{ and }B_2\text{ det.}\big)\Big|
  \le2,
\end{multline}
which is violated by the quantum-mechanical predictions as soon as the
efficiency is nonzero.

There is a third assumption that can be made: that the subset of
observed events does not vary with the settings \citep{Larsson1998,
  Berry2010}. If this subset remains constant, the addition in
equation (\ref{eq:17}) can be performed, and the inequality
\eqref{eq:32} again holds. These three different assumptions are
distinct and none of the three implies the others. Comparing all
three, the no-enhancement assumption applies pointwise in the sample
space, the assumption of constant set of detection applies on the
level of events (subsets in sample space), and the fair sampling
assumption applies on the output statistics.  Pointwise or event-wise
assumptions in the sample space may be thought to be more restrictive
than the latter statistical assumption, but the difference between
these three is small. In recent experiments, the fair sampling
assumption seems to be the most popular.

\subsection{Lower efficiency bounds}
\label{sec:lower-effic-bounds}

Since 82.83\% efficiency is very difficult to achieve in photonic
experiments, improvements are needed.  One important improvement was
found in \citet{Eberhard1993} who rederived the CH inequality
\eqref{eq:28} for detection counts, now sometimes called the Eberhard
inequality.  His approach was to assume that the detection efficiency
is known (constant, and independent detections), and from that
optimize the quantum state to maximize the violation. It turns out
that the maximal violation at non-ideal detection probability occurs
for a non-maximally entangled state. Such a violation can be obtained
as soon as the efficiency is above $2/3 \approx 66.67\%$
\citep{Eberhard1993,Larsson2001}. One example of a state that violates
the CH inequality at efficiency lower than 82.83\% is the one used in
the ``Hardy paradox'' \citep{Hardy1993}.  As the efficiency tends to
2/3, the optimal state tends to the product state
$|A_1=B_1=-1\rangle$.  The probabilities in the CH inequality concern
$+1$ outcomes, so approaching that product state means that all the
probabilities will approach zero. This will lower the noise tolerance
in the test, so that an experimenter aiming to use a non-maximally
entangled state needs to choose it carefully to balance efficiency
demands against noise tolerance.  Finding the optimal state is a
simple matter of numerical optimization, as in \citet{Eberhard1993},
or solving a fourth-degree polynomial equation \citep{Lima2012}.

Two recent photonic experiments \citep{Giustina2013,Christensen2013}
report violation of the CH inequality, closing the efficiency
loophole. Both estimate their efficiencies to close to 75\%. Still,
both are vulnerable to the locality loophole, but this makes photonic
systems the only system where both the efficiency and the locality
loophole has been closed, in separate experiments
\citep[given][]{Weihs1998}. It should be added that the experiment in
\citet{Giustina2013} is also vulnerable to the memory (setting
prediction) loophole of \autoref{sec:memory}, because the settings are
not chosen randomly, nor switched rapidly. There also remains to
discuss another loophole, see \autoref{sec:coinc-pair-ident} below.

The bound can be improved even further by going to, for example, more
sites. A first example of this is the GHZ paradox
\citep{Greenberger1989} for which the efficiency bound is 75\%
\citep{Larsson1998-1}. The CH inequality can be generalized to many
sites and the corresponding bound tends to 50\% in the infinite limit
\citep{Larsson2001}. Using a different family of inequalities, the
bound approaches 0 for infinitely many sites \citep{Buhrman2003}.
Another alternative is to go to higher dimension, where the bound also
can be made to go to zero \citep{Massar2002}.  Unfortunately, these
extreme bounds are impractical to use, e.g., the approach of
\citeauthor{Massar2002} needs 1600-dimensional states to improve over
the CHSH inequality, and is quite sensitive to noise at that point.  A
more practical example is \citet{Vertesi2010} where a bound of 61.8\%
is reached using four-dimensional systems.

One possibility is to use a setup where the system at one site is
different from that at the second site. An example would be to use an
ion in an ion trap at the first site, giving high detection efficiency
$\eta_A$ there, and a photon at the second site, making it possible to
separate the two sites well but at the cost of low efficiency
$\eta_B$. In this case, the high $\eta_A$ makes the bound lower for
$\eta_B$. If $\eta_A=100\%$, the CH inequality can be violated as soon
as $\eta_B>50\%$, a significant improvement \citep{Cabello2007}. This
can be improved further by using more settings, e.g., for the
$I_{3322}$ inequality \citep{Collins2004} a violation is obtained as
soon as $\eta_B>43\%$ \citep{Brunner2007}.  A final interesting
variation is to allow different efficiency between the $\pm1$ channels
at each local site \citep{Garbarino2010}.  This analysis makes it
possible to use more exotic systems like strangeness measurements on
kaons, which may enable high efficiency setups.

There are many alternatives to improve the bounds by choosing system
properly, but it seems that currently low-dimensional systems with few
sites are the most popular, both symmetric and asymmetric with
nonmaximal entanglement. This is possibly because entanglement
generation can be performed efficiently at high rate, giving a high
enough violation in these systems.

\subsection{The coincidence loophole, and fair coincidences}
\label{sec:coinc-pair-ident}

We now turn to another reason that events are missing from the data
used to calculate correlations: determining which local events that
form pairs. The problem of pair identification is especially
pronounced in continuously pumped photonic experiments, but is in
principle present in all experiments that have rapid repetition in the
same physical system.  Relative timing is often used to identify
pairs: if two detections are close in time they are ``coincident,''
otherwise they are not.  The obtained correlations are again
conditional correlations, here conditioned on coincidence,
\begin{equation}
  \label{eq:16}
  E\big(A_iB_j\big|\text{coinc.\ for }A_i\text{ and }B_j\big).
\end{equation}
And again, conditional correlations do not add if conditioning over
different subsets. This results in the ``coincidence'' loophole first
discussed in \citet{Larsson2004}. In that paper a simple local realist
model with varying delays is constructed, that has single-particle
efficiency of 100\% and a conditional coincidence probability of
87.87\%, but still gives the quantum value $2\sqrt2$ in the CHSH
inequality, when conditioning on coincidence. This is quite
remarkable, given the general validity of the 82.84\% bound of
inequality \eqref{eq:26}.

The reason for the high coincidence probability is that the
coincidence loophole concerns a joint property: if two given events
form a pair or not. For comparison, the efficiency loophole concerns a
local property: if an event occurs or not. The two belong to the same
class of loopholes, because the root of the problem is experimental
runs that do not count in the calculated correlations, but the
coincidence loophole is more difficult to handle because it concerns a
joint property of the two events. This can be made explicit as
follows. Under local realism, detection times are random variables
$T^A_i(\lambda)$ and $T^B_j(\lambda)$ that are real-valued, and only
depend on $\lambda$ and the local setting. A coincidence occurs if
these are close enough, say at a maximum distance $\tau/2$ from each
other (making ``the coincidence window width'' equal $\tau$). In terms
of $\lambda$, this corresponds to the subensemble
\begin{equation}
  \label{eq:36}
  \big\{\lambda:\text{coinc.\ for }A_i(\lambda)\text{ and }B_j(\lambda)\big\}
  =\Big\{\lambda:\big|\,T^A_i(\lambda)-T^B_j(\lambda)\big|\le\frac\tau2\Big\}.
\end{equation}
The subensemble in the efficiency loophole where both detections occur
has the simpler structure
\begin{equation}
  \label{eq:37}
  \big\{\lambda:A_i(\lambda)\text{ det.\ and }B_j(\lambda)\text{ det.}\big\}
  =\Big\{\lambda:A_i(\lambda)\text{ det.}\Big\} \cap
  \Big\{\lambda:B_j(\lambda)\text{ det.}\Big\}.
\end{equation}
The latter set is an intersection of two local parts, while the former
cannot be separated into local parts. Also for the coincidence case,
the underlying model is local, but the process of identifying
coincidences makes the subensemble \eqref{eq:36} depend on both
settings $i$ and $j$ in such a way that it cannot be split into local
parts.

In this situation, one could make a ``fair coincidence'' assumption
that would remove the loophole: (paraphrasing the CHSH criterion) that
if a pair of photons emerges from the beam-splitters, the probability
of them registering as coincident is independent of the settings, or
\begin{equation}
  \label{eq:34}
  P\big(A_i=\pm1\cap B_j=\pm1\cap
  \text{coinc.\ for }A_i\text{ and }B_j\big)
  =c P\big(A_i=\pm1\cap B_j=\pm1\big).
\end{equation}
Again, a possible mental image might be $(A_i,B_j)=(\pm1,\pm1)$
denoting coincident emergence from an ideal polarizing beamsplitter,
and subsequent delays in the detectors. Using the fair-coincidence
assumption, and the same reasoning as above, the CHSH inequality
applies to the underlying local realist model for the unobserved ideal
outcomes, so that a CHSH inequality applies for the coincidences,
\begin{multline}
\label{eq:35}
  \Big|E\big(A_1B_1\big|\text{coinc.\ for }A_1\text{ and }B_1\big)
  +E\big(A_1B_2\big|\text{coinc.\ for }A_1\text{ and }B_2\big)\Big|\\
  +\Big|E\big(A_2B_1\big|\text{coinc.\ for }A_2\text{ and }B_1\big)
  -E\big(A_2B_2\big|\text{coinc.\ for }A_2\text{ and }B_2\big)\Big|
  \le 2.
\end{multline}
Also this is violated by the quantum-mechanical predictions as soon as
the efficiency is nonzero.

\subsection{Bounds for the coincidence loophole}
\label{sec:bounds-coinc-looph}

Deriving an inequality that is valid without the fair-coincidence
assumption is more complicated than for the efficiency loophole. Here,
it is not possible to assign an outcome for the missing data, because
this will give a nonlocal assignment. The assigned outcome will depend
on the remote setting, because of the reference to the joint property
\eqref{eq:36} of the two events.  Therefore, another technique needs
to be employed that avoids value assignment to the outcomes. This is
done in \citet{Larsson2004} and the method is similar to that used in
\citet{Larsson1998} that avoids using assigned values for
non-detections.  The result is the generally valid CHSH-like
inequality
\begin{multline}
  \label{eq:38}
  \Big|E\big(A_1B_1\big|\text{coinc.\ for }A_1\text{ and }B_1\big)
  +E\big(A_1B_2\big|\text{coinc.\ for }A_1\text{ and }B_2\big)\Big|\\
  +\Big|E\big(A_2B_1\big|\text{coinc.\ for }A_2\text{ and }B_1\big)
  -E\big(A_2B_2\big|\text{coinc.\ for }A_2\text{ and }B_2\big)\Big|
  \le \frac6\gamma-4.
\end{multline}
where
\begin{equation}
\label{eq:39}
  \gamma=\min_{i,j} P\big(\text{coinc. for }A_i\text{ and }B_j\big).
\end{equation}
This probability is not so easy to estimate from experimental data,
but it has been conjectured \citep{Larsson2004} that the same bound
holds for the relevant conditional coincidence efficiency
\begin{equation}
  \label{eq:40}
  \eta=\min_{i,j} \Big\{P\big(\text{coinc. for }
  A_i\text{ and }B_j\big|B_j\text{ det.}\big),\
  P\big(\text{coinc. for } A_i\text{ and }B_j\big|A_i\text{ det.}\big)\Big\}.
\end{equation}
This is a measure of the ``apparent efficiency,'' because it can be
less than 1 even if the detectors have ideal efficiency. Again, it is
important that the apparent efficiency of the entire experiment is
taken into account.

Also here, an event-ready setup can be used to remove the difficulty
of estimating~$\gamma$. In such a setup, coincidence is determined by
closeness to the event-ready indication time, not by small time
difference to the remote detection. The result is well-defined
experimental runs that are independent of the measurement settings,
similar to the detector-efficiency case. Then, event-ready signals
from the source that are not coincident with detections at the
measurement sites can be interpreted as missing detections, locally
assigned some value (say 0), and included in a
detector-efficiency-like analysis.  In this situation, we recover
\begin{multline}
  \label{eq:41}
  \Big|E\big(A_1B_1\big|\text{coinc.\ for }A_1\text{ and }B_1\big)
  +E\big(A_1B_2\big|\text{coinc.\ for }A_1\text{ and }B_2\big)\Big|\\
  +\Big|E\big(A_2B_1\big|\text{coinc.\ for }A_2\text{ and }B_1\big)
  -E\big(A_2B_2\big|\text{coinc.\ for }A_2\text{ and }B_2\big)\Big|
  \le \frac4\eta-2.
\end{multline}

Another simpler option is to use fixed time slots. This is natural in
a pulsed-pump experiment, but also possible in a continuously pumped
experiment. The important property is well-defined experimental runs
that are independent of the measurement settings.  Then, coincidence
is determined by presence in the same time-slot, not by small time
difference to the remote detection.  Slots that are empty can locally
be assigned some value (say 0), and the system can be analyzed in a
detector-efficiency-like manner \citep{Larsson2013}.  Note that there
actually is something to prove here, to re-establish \eqref{eq:41}:
that delays \emph{on both sides} do not influence the bound
negatively, raising it.  A CH-like inequality that includes
restriction to coincidences can also be derived by this method, and it
reads
\begin{equation}
  \label{eq:42}
  \begin{split}
    & P(A_1=B_1=1\cap\text{coinc. for }A_1\text{ and }B_1)\\
    &+P(A_1=B_2=1\cap\text{coinc. for }A_1\text{ and }B_2)\\
    &+P(A_2=B_1=1\cap\text{coinc. for }A_2\text{ and }B_1)\\
    &-P(A_2=B_2=1\cap\text{coinc. for }A_2\text{ and }B_2)\\
    &-P(A_1=1)-P(B_1=1) \le 0.
  \end{split}
\end{equation}

The third and final method to avoid using the fair-coincidence
assumption is to use different-sized time windows for the different
coincidence probabilities. This method is most easily applied on
inequalities that contain probabilities directly, since it is quite
complicated to handle this in correlation-based inequalities. For
example, in the CH inequality the time windows can be arranged as
follows:
\begin{equation}
  \begin{split}
    \label{eq:43}
    &\text{coinc. for }A_1\text{ and }
    B_1:\ \big|\,T^A_1-T^B_1\big|\le\frac\tau2,\\
    &\text{coinc. for }A_1\text{ and }
    B_2:\ \big|\,T^A_1-T^B_2\big|\le\frac\tau2,\\
    &\text{coinc. for }A_2\text{ and }
    B_1:\ \big|\,T^A_2-T^B_1\big|\le\frac\tau2,\\
    &\text{coinc. for }A_2\text{ and }
    B_2:\ \big|\,T^A_2-T^B_2\big|\le\frac{3\tau}2.
  \end{split}
\end{equation}
Then, the subensemble that gives coincidences for all three of
$A_1B_1$, $A_1B_2$, and $A_2B_1$ will also give a coincidence for
$A_2B_2$. This will also enable inequality \eqref{eq:42}.

A few recent experiment address the coincidence loophole explicitly,
but because of inequality \eqref{eq:41}, all performed pulsed-pump
optical experiments do this implicitly: they are not vulnerable to the
coincidence loophole, but only the efficiency loophole.  In
\citet{Aguero2012} a pulsed-pump optical system is used to further
probe four different coincidence properties of local realist models
for their system, but they cannot address the coincidence loophole in
Bell tests as such because of the pulsed pump: their setup is simply
not vulnerable to the loophole. A pulsed-pump setup is used in
\citet{Christensen2013}, so that it is free of the coincidence
loophole, in addition to the efficiency loophole. Finally, the
continuously pumped optical experiment by \citet{Giustina2013} is free
of the efficiency loophole, and also of the coincidence loophole as
shown in \citet{Larsson2013}.

\subsection{The postselection loophole, and realism assumptions}
\label{sec:posts-auxil-real}

Both the efficiency loophole and the coincidence loophole can be
viewed as caused by deficiencies in the experimental equipment. Low
overall experimental efficiency or coincidence probability can in
principle be counteracted by improving the experimental setup. We will
now turn to a different setup, for which the problem is built in at a
deeper level and cannot so easily be removed.

\begin{figure}[t]
  \centering
  \includegraphics{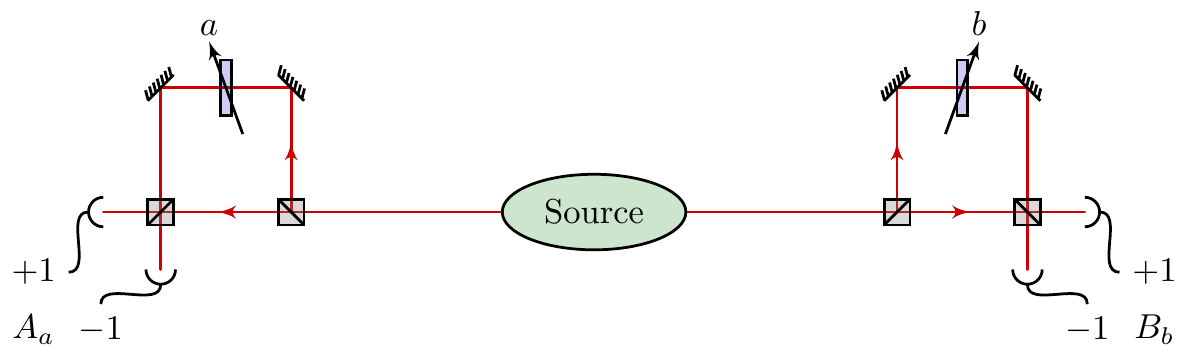}
  \caption{The Franson interferometer. The source sends out
    time-correlated photons at unknown moments in time. These travel
    through unbalanced (but equal) Mach-Zehnder interferometers with
    variable phase delays $a$ and $b$. If the detections are
    coincident and $a+b=0$, then $A_a=B_b$ with probability
    one.}
  \label{fig:2}
\end{figure}

The experiment in question is the Franson interferometer
(\cite*{Franson1989}), see \autoref{fig:2}. The setup uses a source
that emits time-correlated photons at unknown moments in time, and two
unbalanced Mach-Zehnder interferometers. The interferometers should
have a path difference that is large enough to prohibit first-order
interference, which means that the difference must be larger than the
coherence time of the photons emitted from the source.  When this is
the case, the probability is equal for a photon to emerge from each
port of the final beamsplitter.

Now, the interferometer path-differences should be as equal as
possible; the path-difference difference (repetition intended) should
be small enough to make photon pairs where both photons ``take the
long path'' and both photons ``take the short path''
indistinguishable.  Quotation marks are used here to remind the reader
that photons are not particles, but quantum objects and as such, do
not take a specific path. Then, there are two possibilities for the
photons to reach the detectors, and these interfere at the second
beamsplitter.

If the path-difference difference is smaller than the coherence length
of the pump photons, the phase difference between the two
wave-functions that meet on the second beam splitter is stable. And if
the total phase delay $a+b=0$, then a photon emerging in the $+1$
channel on the left is always accompanied by a photon emerging in the
$+1$ channel on the right, and the same for the $-1$ channels.  As a
function of the total phase delay, we have
\begin{equation}
  \label{eq:44}
  \big\langle A_aB_b\big|\text{coinc.\ for }A_a\text{ and
  }B_b\big\rangle=\cos(a+b).
\end{equation}
Thus, the interference is not visible as a change in output intensity,
as in first-order interference, but instead as a sinusoidal
\emph{correlation} of the outputs. And a sinusoidal correlation
immediately points to violation of the CHSH inequality.

If the photons are not coincident, the delays are different. Then one
photon took the long path and the other took the short, and the
emission time can be calculated easily as the early detection time
minus the short path length divided by the speed of light. In this
case, there is only one possible way that the photons could have
reached the detectors, and there will be no interference.  Therefore,
these events are not useful in our setting and are discarded.

And this is precisely the problem: We postselect events based on
coincidence, as in the coincidence loophole above. But here, the
selection is built into the setup, rather than a property of the
equipment. It is there even with ideal equipment. And indeed, even
though the correlation is sinusoidal and seems to violate the CHSH
inequality, it does not. Even with a pulsed source at the appropriate
rate, the relevant inequality with conditioning on coincidences is
\eqref{eq:41} with $\eta=50\%$ which means that it reads
\begin{multline}
  \label{eq:45}
  \Big|E\big(A_1B_1\big|\text{coinc.\ for }A_1\text{ and }B_1\big)
  +E\big(A_1B_2\big|\text{coinc.\ for }A_1\text{ and }B_2\big)\Big|\\
  +\Big|E\big(A_2B_1\big|\text{coinc.\ for }A_2\text{ and }B_1\big)
  -E\big(A_2B_2\big|\text{coinc.\ for }A_2\text{ and }B_2\big)\Big|
  \le 6,
\end{multline}
which is not a useful bound.  Indeed, there exists a local realist
model that gives the same outputs as quantum mechanics predicts
\citep{Aerts1999}, for ideal experimental equipment, at all values of
$a$ and $b$. The root of the problem is the built-in postselection;
this is called the ``postselection'' loophole, or sometimes the
geometric loophole.

As with the detection and coincidence loopholes this can be avoided in
a number of ways. The first is to assume that the photon really does
take one path through the interferometer (\cite{Franson1999};
\cite*{Franson2009}). This assumption of ``path realism'' implies that
the photon will travel either through the long path, or through the
short path, as decided at the first beamsplitter. The decision needs
to be independent of the phase-delay setting to ensure that the
subensembles are the same, so that the conditional correlations can be
added directly. This independence could be part of the assumption, but
it is better to arrange the experiment so that independence is ensured
by locality. Then, the choice of path and the choice of phase-delay
setting must be space-like separated, and to enforce this the
experimenter needs a large distance between the beamsplitter and the
phase-delay internally in each interferometer. In short, he or she
needs to enforce locality inside the interferometer, instead of
between measurement sites and the source, as in the standard Bell
setup. If this is done and path realism is assumed, one can
re-establish inequality \eqref{eq:35} with the bound 2, and a quantum
violation.

A second alternative is to use the elements of reality that are at
hand. These are: the measurement outcomes $\pm1$ and the time of
emission. The latter is an EPR element of reality because the entire
interferometer can be removed at one site without affecting the other
site, and the emission time can be calculated from the detection time
there. This means that the emission time of the other photon of the
pair can be predicted without disturbing it. Furthermore, the delay
must be specified by the local realist model \citep{Aerts1999}, and
there are two equally large subensembles inside the coincidence
subensemble: early-early events, and late-late events. The bound for
the early-early events cannot be lowered, but the late-late bound for
the CHSH inequality is 2 if the system is set up with some care. In
short, the phase-delay needs to be as close to the detectors as
possible; for details see \citet{Aerts1999,Jogenfors2014}.  This needs
no auxiliary assumptions and no space-like separation of beam-splitter
and phase delay in the same interferometer.  Since the early-early and
late-late events occur with the same probability, we arrive at
\begin{multline}
  \label{eq:46}
  \Big|E\big(A_1B_1\big|\text{coinc.\ for }A_1\text{ and }B_1\big)
  +E\big(A_1B_2\big|\text{coinc.\ for }A_1\text{ and }B_2\big)\Big|\\
  +\Big|E\big(A_2B_1\big|\text{coinc.\ for }A_2\text{ and }B_1\big)
  -E\big(A_2B_2\big|\text{coinc.\ for }A_2\text{ and }B_2\big)\Big|
  \le 3,
\end{multline}
which is better than inequality \eqref{eq:45}, but not good enough.
Fortunately, \citet{Pearle1970} and \citet{Braunstein1990} come to our
rescue. These papers contain a class of generalized Bell inequalities
known as ``chained'' Bell inequalities, with more terms, e.g., the
six-term inequality
\begin{multline}
\label{eq:47}
  \Big|E\big(A_1B_1\big)+E\big(A_1B_2\big)\Big|
  +\Big|E\big(A_2B_2\big)+E\big(A_2B_3\big)\Big|
  +\Big|E\big(A_3B_3\big)-E\big(A_3B_1\big)\Big|
  \le 4.
\end{multline}
Using this for the late-late subensemble and the trivial bound 6 for
the early-early, we obtain
\begin{equation}
\label{eq:48}
  \begin{split}
    &\Big|E\big(A_1B_1\big|\text{coinc.\ for }A_1\text{ and }B_1\big)
    +E\big(A_1B_2\big|\text{coinc.\ for }A_1\text{ and }B_2\big)\Big|\\
    &\qquad+\Big|E\big(A_2B_2\big|\text{coinc.\ for }A_2\text{ and }B_2\big)
    +E\big(A_2B_3\big|\text{coinc.\ for }A_2\text{ and }B_3\big)\Big|\\
    &\qquad+\Big|E\big(A_3B_3\big|\text{coinc.\ for }A_3\text{ and }B_3\big)
    -E\big(A_3B_1\big|\text{coinc.\ for }A_3\text{ and }B_1\big)\Big|
    \le 5.
  \end{split}
\end{equation}
The quantum-mechanical prediction is $6\cos\pi/6=5.196$, so this gives
a violation. The above inequality is more sensitive to noise than the
usual Bell inequality; this is the price one has to pay for avoiding
the assumption of path realism. Adding more terms in the chain will
decrease the noise sensitivity but the minimum is reached already at
ten terms \citep{Jogenfors2014}.

A third alternative method to remove the loophole is to change the
experiment setup so that path really is an EPR element of
reality. There have been three proposals of how to do this. The first
by \citet{Strekalov1996} uses polarization-entangled photons and a
polarizing beamsplitter to force the two paths to be the same: the
only alternatives are long-long and short-short pairs, which means all
pairs will be coincident. There is no postselection anymore, so of
course, there is no postselection loophole (but other loopholes will
still apply).

A second proposal by \citet{Brendel1999} is to use a pulsed source
with a similar unbalanced Mach-Zehnder interferometer preceding the
pump, and moving mirrors in place of the first beamsplitters at the
analyzer stations. This ensures that photon pairs created by the
non-delayed pump pulse take the long path at the analyzers, and photon
pairs created by the delayed pump pulse take the short path at the
analyzers. The effect is the same as for the polarization scheme
above, that all pairs will be coincident. There is no postselection
anymore, so consequently, there is no postselection loophole (but
other loopholes will still apply).

This is known as time-bin entanglement, but there is one caveat here.
Standard time-bin entanglement experiments do not use active mirrors,
but instead ordinary beamsplitters, so that the analyzer stations are
actually exactly the same as in Franson's scheme. This is suggested
already in \citet{Brendel1999} which only mentions a 50\% loss as
detrimental effect, which is true if the loss occurs at the source;
then there is no adverse effects besides the loss.  However, at the
analyzer stations, the effect is random delays in the detections of
the photons similar to the original Franson proposal, meaning that the
50\% loss is due to postselection and this re-enables the
postselection loophole.  Experiments that want to be loophole-free
must use active moving mirrors, and would constitute ``genuine''
time-bin entanglement.

The third and final method to ensure that path is an EPR element of
reality was proposed in \citet{Cabello2009-1}, and the suggestion is
to interchange the long paths of the analyzer stations. Then, the long
path from the first beam-splitter is directed towards the opposite
station. These first beamsplitters should now be considered part of
the source and located at the source. With this configuration, path is
an EPR element of reality because it can be remotely predicted from
the result of a local measurement. If only one photon emerges from the
paths, the other photon of the pair can be predicted to emerge from
the corresponding path at the remote station, giving a coincidence and
coincident events will behave as in the Franson setup, giving
interference. A pair for which there is no coincidence will behave
differently: both photons of the pair will appear at one of the
analyzer stations, and can be discarded by an entirely local
process. There is postselection, but no loophole; this setup gives
``genuine'' energy-time entanglement (but other loopholes will still
apply). An experimental violation of the CHSH inequality in this
system was reported in \citet{Lima2010} at a distance of 1 m, while
non-maximal entanglement was created in \citet{Vallone2011}. Finally,
in \citet{Cuevas2013} one of the interferometers had 1 km long arms,
and was still stabilized well enough to give a violation of the CHSH
inequality.

\section{Conclusions}
\label{sec:conclusions}

There are a number of loopholes that may occur in experimental tests
of Bell inequalities, but almost all the loopholes can be handled.
General advice for Bell-inequality-violating experimenters would be
the following.
\begin{description}
\item[\quad Perfect correlation.] Reduce noise as far as possible.
  Perfect correlation is not needed, but there must be a decent-sized
  violation (\autoref{sec:determ-vs.-stoch}).
\item[\quad Simplifying assumptions.] Avoid assuming symmetry of the
  setup; clearly state assumptions that cannot be avoided. Make sure
  that outcomes that occur in the inequality are measured
  individually. Do not remove accidentals
  (\autoref{sec:viol-from-quant}).
\item[\quad Finite sample.] Use proper statistics, a large trial, and
  be clear on what statistical assumptions are used.  It is preferable
  to do a proper hypothesis test to just reporting the violation as a
  number of standard deviations (\autoref{sec:exper-viol-finite}).
\end{description}
Locality-related loopholes should should be avoided as follows.
\begin{description}
\item[\quad Fast switching.] Make sure that no signal about the local
  choice of setting can reach the remote site before the measurement
  has finished (\autoref{sec:local-fast-switch}).
\item[\quad Memory.] Use a good source of randomness for the switching
  so that the settings cannot be predicted. If independent trials is
  assumed, state that; alternatively adjust the statistical analysis
  (\autoref{sec:memory}).
\item[\quad Freedom of choice.] Make sure that the source and the
  local choice of settings are independent of each other
  (\autoref{sec:freed-choice-superd}).
\end{description}
There is one loophole that cannot be avoided:
\begin{description}
\item[\quad Superdeterminism.] This cannot be closed by scientific
  methods, but should instead be ruled out on philosophical grounds;
  the basic mode of operation of a scientist is to discover the laws
  of nature by experimentation, which is not possible under
  superdeterminism.  Therefore, a natural choice is to proceed under
  the assumption that superdeterminism does not hold
  (\autoref{sec:freed-choice-superd}).
\end{description}
Finally, efficiency-related loopholes should also be avoided.
\begin{description}
\item[\quad Ideal efficiency.] The efficiency loophole can be avoided
  altogether by using a system that gives outcomes for every
  experimental run (\autoref{sec:efficiency}). If such a system is not
  used, see below.
\item[\quad Fair sampling, or no-enhancement.] State clearly if one of
  these assumptions is used. Avoid them in loophole-free experiments
  (\autoref{sec:efficiency-sampling}; \ref{sec:prob-not-corr}).
\item[\quad Non-ideal efficiency.] Use high-efficiency equipment, and
  report the efficiency of the whole setup.  Be clear on how the
  efficiency is estimated (\autoref{sec:non-detection-outcomes}).
\item[\quad Event-ready setup.] Use this to assign values (e.g., 0) to
  missing outcomes. This eliminates the efficiency loophole, but will
  lower the violation (\autoref{sec:non-detection-outcomes}).
\item[\quad Lower efficiency bounds.] Use the Clauser-Horne inequality
  with nonmaximal entanglement, asymmetric systems, or
  higher-dimensional systems, several sites, and so on.
  (\autoref{sec:lower-effic-bounds}).
\item[\quad Coincidences.] Be clear on how pairs are identified, and
  report the apparent coincidence efficiency of the whole setup. An
  event-ready setup avoids the problem, while pulsed pump or
  appropriate time windows reduces it to the simpler efficiency
  loophole (\autoref{sec:coinc-pair-ident};
  \ref{sec:bounds-coinc-looph}).
\item[\quad Postselection.] Avoid postselection, and if this is not
  possible, clearly state what assumptions eliminate the loophole.
  Alternatively, use a setup that eliminates the loophole
  (\autoref{sec:posts-auxil-real}).
\end{description}
In general, assumptions other than local realism should be absent from
a loophole-free experiment; any and all extra assumptions should be
stated clearly.

These considerations do have implications outside tests of local
realism. For device-independent security of Bell-inequality based
quantum cryptography \citep{Acin2007}, it is crucial to rule out
alternative local realist descriptions of the key-generation process.
If there are loopholes that allows a local realist model to explain
the received bits, it is possible that these are generated by a faked
system mimicking quantum violations through carefully constructed
classical systems \citep{Larsson2002-1} or by controlling the detector
\citep{Gerhardt2011}.  Device-independent security crucially relies on
a loophole-free experiment.

It may seem pedantic to insist on removing loopholes from Bell
inequality tests, certainly when no other generally accepted
explanation of the observed phenomena exists. But, arguably, the point
of a loophole-free Bell test is to \emph{definitively} rule out any
local realist theory. And this is not possible as long as additional
assumptions about the system are needed along with those of local
realism, as long as the experiments are vulnerable to loopholes.

\printbibliography

\end{document}